# Facile patterning of functional materials via gas-phase 3D printing

*Cesar Arturo Masse de la Huerta,‡ Viet H. Nguyen,‡ Abderrahime Sekkat,‡ Chiara Crivello,‡ Fidel Toldra-Reig, Pedro Veiga, Carmen Jimenez, Serge Quessada, David Muñoz-Rojas\**

Université Grenoble Alpes, CNRS, Grenoble INP, LMGP, Grenoble, France

‡ *These auhors contributed equally to this work*
E-mail: david.munoz-rojas@grenoble-inp.fr

**Abstract**

Spatial Atomic Layer Deposition (SALD) is a recent approach that is up to two orders of magnitude faster than conventional ALD, and that can be performed at atmospheric pressure and even in the open air. Previous works have exploited these assets to focus on the possibility of high-rate, large-area deposition for scaling up into mass production. Conversely, here we show that SALD indeed represents an ideal platform for the selective deposition of functional materials by proper design and miniaturization of SALD close-proximity heads. In particular, we have used the potential offered by 3D printing to fabricate custom close-proximity SALD injection heads. By using 3D printing, the heads can be easily designed and readily modified to obtain different deposition areas, free-form patterns, and even complex multimaterial structures. The heads can be printed in different materials to adjust to the chemistry of the precursors and the deposition conditions used. Polymeric heads can be used as cheap (even disposable) heads that are both used for performing deposition and for prototyping and optimization purposes. Finally, by designing a miniaturized head with circular concentric gas channels, 3D printing of functional materials can be performed with nanometric resolution in Z. This constitutes a new 3D printing approach based on gaseous precursors. Because the selective deposition strategies presented here are based on the SALD process, conformal and continuous thin films of functional materials can be printed at low temperatures and with high deposition rate in the open air. Our approach represents a new versatile way of printing functional materials and devices with spatial and topological control, thus extending the potential of SALD and ALD in general, and opening a new avenue in the field of area-selective deposition of functional materials.

## 1.- Introduction

The possibility to pattern functional materials in a simple and cheap way is of key importance for many applications and technologies.[1–4] At the macro level, direct printing of functional materials is possible with different approaches, one of the most popular being inkjet printing.[5]

Concerning the micro and nano level, while standard approaches based on lithography are effective and mature, they involve the design and fabrication of masks, many steps and expensive equipment. Furthermore lithography and patterning are often performed in cleanrooms, thus adding to the cost of the final devices. In this context, there have been much progress in bringing the resolution of inkject printing down to the micrometer level.[6] And other new bottom-up strategies are being developed, for example based on microfluidics.[7] Another new approach to so-called area-selective deposition (ASD) has been developed in the last years in the field of atomic layer deposition (ALD).[3,8] In ALD, solid-gas, surface-limited, self-terminating reactions take place.[9–11] This results in very compact and continuous thin films with a sub-nanometer thickness control and unique conformality, key assets for engineering the interfaces and surfaces of different functional materials and devices.[12–15] Given the surface–limited nature of ALD, several strategies can be implemented to control the reactivity of ALD precursors towards different surfaces to generate selectivity and result in ASD. These involve the use of a substrate with different materials (growth inhibition or delay taking place in one of them) or the use of self-assembled monolayers to block the growth in certain parts of the surface, as recently reviewed by Mackus et al.[16] While these approaches are very appealing and have proven to be efficient, there is still the need to have a surface with different materials, and in most cases a pre-patterning step is necessary. Also, some of the ASD approaches require intermediate etching steps.[17] Finally, these approaches suffer from the inherent low deposition rate of ALD and the use of expensive vacuum equipment.

In the last years, Spatial ALD (SALD) has established itself as a high-throughput, low-cost alternative to conventional ALD. SALD is based on the spatial separation of precursors, as opposed to the sequential (temporal) separation in ALD.[18,19] Thus, precursors are continuously injected in different locations, which are spatially separated by an inert gas region. As the sample is exposed to the different regions, the standard ALD cycle is reproduced. Because no purge steps are needed, SALD can be up to two orders of magnitude faster than ALD.[20,21] In addition, SALD can be performed at atmospheric pressure (AP-SALD), thus resulting more convenient for scaling-up than vacuum-based ALD.[22–24] But because SALD is based on the same surface-limited, self-terminating chemistry than ALD, the low temperature processing, unique conformality and thickness control to the sub-nanometer scale of ALD are maintained in SALD.[25] The quality of thin films deposited by SALD has been shown through their intense application to photovoltaics.[26,27]

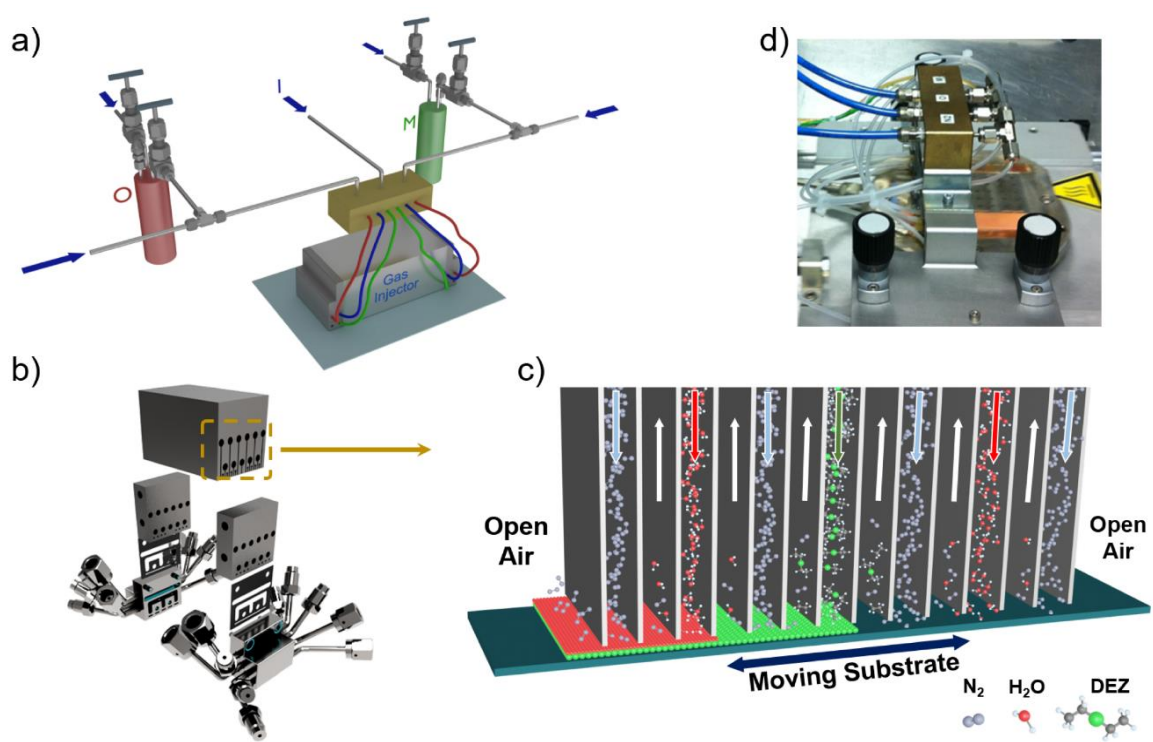

**Figure 1.** a) Close-proximity SALD system based on a gas manifold injector working in the open air (no deposition chamber used). The precursors are stored in bubblers through which an inert gas (Ar or $N_2$) is bubbled to carry out the precursor volatile molecules. b) 3D scheme of the different parts in a typical close-proximity SALD head. The different flows carrying the precursors and inert gas are separated in parallel channels. Several pieces and connections are needed to distribute the different flows through the respective channels. c) Close up of the bottom of the deposition head on top of a substrate. Precursors are injected through their respective channels and then exhausted through adjacent exhaust channels, thus being exposed to a moving substrate, resulting in the reproduction of the typical ALD cycles. The scheme shows the deposition of ZnO from DEZ (Diethyl zinc) and water. d) Image of a conventional head installed in our SALD system, showing the complex set of pipes and connections.

While AP-SALD can be implemented in many different ways,[19,28] the close-proximity approach initially developed by Levi et al. is particularly interesting (**Figure 1**a).[29] It is based on a manifold injection head (Figure 1b) in which the different precursors and inert gas are distributed along parallel slots over a moving substrate. By placing the substrate close enough to the surface of the substrate (typically in the range of hundreds micrometers), the crosstalk between precursors is prevented and only gas-solid reactions on the substrate typical of ALD can take place (Figure 1c). As opposed to other SALD approaches, no reaction chamber is used and the process takes place in the open air, thus effectively resembling a functional materials printer.

The deposition manifold head is the key piece of such SALD systems, and thus could in principle be easily adapted to the size of the sample to scale up (for large-area deposition) or down (for ASD or patterning) the process. The deposition head has traditionally been fabricated in metal by different approaches. In the original patent by Levi et al. the gas distribution inside the head was done through channels that were obtained by putting together a series of plates with cut channels or engraved grooves.[30] Such a head is not ideal since it requires disassembly, reassembly and polishing every time a channel gets blocked with materials that are not easy to dissolve away. More recently, simpler heads have been obtained by conventional machining approaches, such as water jet cutting, to get the gas distribution channels, followed by welding of different parts (Figure 1b).[31] Nevertheless, in both cases the procedures for fabricating the heads are expensive and complex and/or time consuming. In addition, the use of conventional fabrication approaches limits the flexibility and complexity of the design of the head. Finally, many connections and pipes are necessary to distribute the different flows, increasing the likelihood of having leaks or contamination (Figure 1d).

3D printing (aka additive manufacturing) is an emerging fabrication technique that allows the production of freeform parts without the need for machining or casting.[32,33] Initially developed for prototyping purposes, 3D printing is a layer-by-layer fabrication method that creates structures from a computer-aided design (CAD). The technique is gaining momentum in the last years, and is already used by a number of institutions and companies including NASA, Airbus or Boeing, the later having obtained the Federal Aviation Administration (FAA) certification for the first 3D printed metallic parts.[34] Since its first development in 1984,[35] it is now possible to print from basic polymeric materials at the consumer level to metal parts at the laboratory and industrial level.[36] The last standard materials joining the game are ceramics[37], glass [38] and even Teflon. Several 3D printing approaches have been developed over the years that are indeed associated with the type of material being printed.[32,39,40] 3D printing is thus a magician box offering almost unlimited possibilities for the design of customized complex pieces made of many different materials, from metals to chemically resistant polymers. The great potential of 3D printing has also been exploited in the last years for the deposition of advanced functional materials, such as hydrogels, hybrid materials, porous silica or quartz, sugar scaffolds or biomaterials, among others.[41–48]

In this work, we combine the potential of 3D printing with the unique assets of SALD to present a new approach for the printing of functional materials. In particular, we apply 3D printing to fabricate customized monolithic close-proximity heads for our open-air SALD system. The obtained heads are fabricated in a matter of hours. They are as well much cheaper and easier to tune than standard heads fabricated by conventional approaches. We have used a low-cost table top 3D printer to fabricate plastic heads with different shapes and sizes. For example, 3D printing allows the integration of the whole gas distribution system inside the body of the head, simplifying the connection of the head. Due to their low cost, they can be considered as disposable components since they can be easily fabricated if blocked or damaged during deposition. When needed, heads can also be printed in metal or other materials. We show that by simple modifications of the head designs, we can easily adapt to different substrate sizes and perform ASD and patterns on a surface without the need of previous patterning or masking steps. Finally, by combining a 3D printed miniaturized SALD head with a XYZ table, freeform motives and gas-phase additive manufacturing of functional materials can be realized. Such a simple open-air approach for the patterning and 3D printing of functional materials with high conformality and nanometer resolution in the Z axis offers a great potential for many applications.

## 2.- Results and discussion

As illustrated in **Figure 2**, customized heads are designed by 3D CAD approaches. The shape and dimensions of the different channels can be previously optimized by computational fluid dynamics (CFD) simulations. Figure 2a shows the 3D scheme of the inner distribution channels for the different gas including reactive precursors and inert gas for a monolithic, custom-designed head. The possibility to 3D print the heads allows the use of such curved, bell-like shaped patterns which are more indicated than the rectangular ones used in our conventional head to obtain a homogeneous flow exiting the head. The scheme also shows how the distribution of the three main flows, i.e. metallic precursor, co-reactant (water or an oxidizer for the deposition of oxides) and inert gas, are easily distributed inside the head along the different channels. Figure 2b shows a head just after being printed in the Formlabs clear resin, incorporating the distribution channels shown in the design in Figure 2a. (The different inner channels with the shapes are clearly seen). As opposed to the complex head shown in Figure 1, 3D printing thus allows the fabrication of monolithic heads in which the gas distribution is implemented and where the shape of the channels can be easily modified. Figure 2c shows the printed head to which some fittings have been added so that it is ready to be used in the SALD system by simply connecting the 3 flow lines and the exhaust line. As shown, 3D printing allows to use custom-designed channel shapes and configurations. It is thus very easy to modify and optimize them, to change the number of metallic outlets in a head, or to selectively collect the gasses from particular exhaust channels (as is the case in the head shown in Figure 1a-c), for example for *in situ* characterization[49] or for optimizing the process.

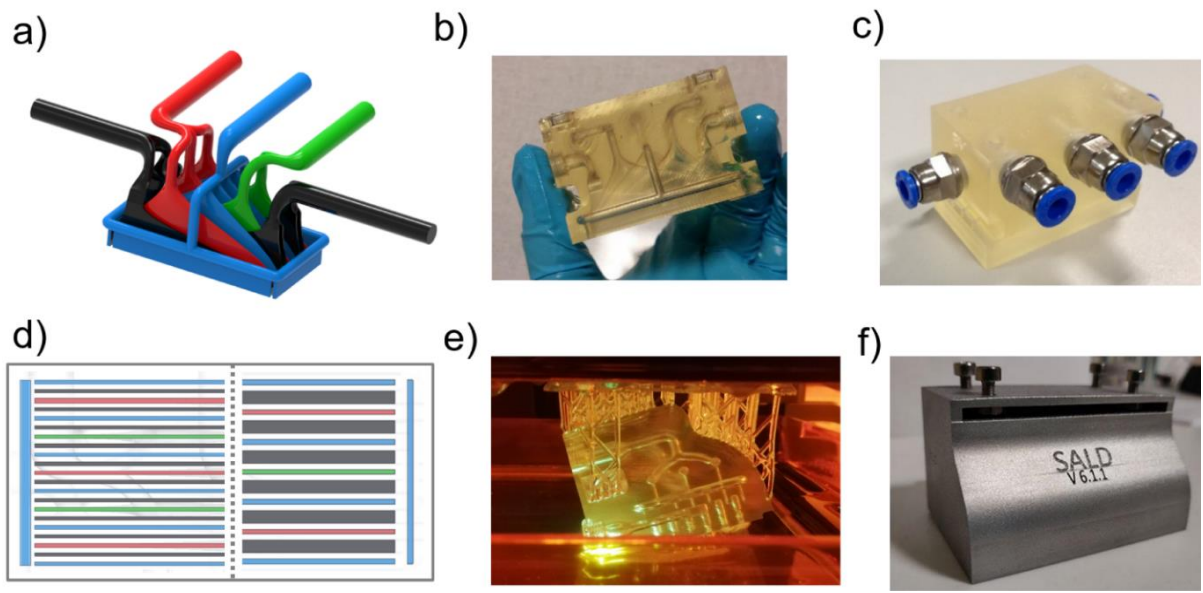

**Figure 2.** a) 3D scheme of the inner distribution channels for the different gases: metallic precursor in green, co-reactant in red, inert gas in blue and exhaust in black b) Head printed with clear resin. Distribution channels can be observed. c) Printed head with fittings added, ready to *plug and play* in the SALD system. d) Bottom view of two different head with different outlet designs. Left, design corresponding the head presented in a). Right, design with only one metal precursor outlet and wider exhaust channels and separation between different outlets. e) In-situ printing of a head in a Form 2 from fromlabs in Clear resin, with the outlet design in the right part of d). Inner channels can be observed. f) The same head design 3D printed in metal.

Indeed, given the nature of close-proximity SALD, it is very important to control the effective separation of the precursors on top of the substrate in order to ensure a proper ALD reaction.[50] This is normally controlled by adjusting the gap between the head and the substrate and by adjusting the different flows. With the availability of low-cost 3D printed heads, another parameter that can be easily modified and tuned is for example the distance between the different outlets (i.e. the design of the head), which directly affects the exposure time of precursors to the substrate surface. Figure 2d shows the bottom of two different head designs. To the left, the outlets in the design from Figure 2a are shown. To the right, the outlets correspond to a design in which only one precursor channel (green) has been included, the exhausts (grey) are wider and the separation between the different outlets is bigger. This is convenient since the larger separation prevents precursor cross-talk more effectively, and thus the proper parallel alignment of the head with the substrate and the value of the gap are less important, facilitating the operation of the system and improving reproducibility. However, if both the left and right designs are optimized, the left part of the head will result in twice faster growth rate compare to the right one because of the difference in number of precursor channels. Figure 2e shows a picture of this alternative head while being printed in clear resin, where the inner conducts are clearly seen. Plastic heads are thus very convenient for prototyping and optimization, and as a cheap, even disposable, alternative to conventional heads. But for depositions at high temperatures (>200 °C), or when chemical incompatibility may exist with the precursors, the use of plastic heads could be an issue. In this case, and thanks to the advancement of the different 3D printing technologies, heads with elaborate inner patterns such as the ones shown here can also be printed in tougher and more thermally stable materials, namely, ceramics or metals. Figure 2f shows a picture of 3D printed metallic head with the same design as in Figure 2e. Being able to print heads with elaborate patterns in different materials thus extends the versatility of this simple approach.

**Figure 3** shows other examples of the possibilities offered by implementing the 3D printing of SALD heads. Figure 3a shows two versions of the head shown in Figure 2e having different outlet lengths, thus yielding a smaller deposition area. Indeed, when using the standard, non-easily modifiable head, the deposition area stays always the same, regardless of the size of the sample to be coated. This has two main drawbacks. On one hand, there is a lot of precursor wastage when depositing on samples smaller than the deposition head. On the other hand, it is virtually impossible to have a continuous uniform film by doing sequential coating on different parts of the substrate for samples larger than the head. Thus, being able to easily adjust the size of the outlets to the sample to be coated is very interesting. Figure 3a shows $Cu_2O$ thin films deposited using the head with the standard outlet length (5 cm) and the one made with 2.5 cm long outlets. In both cases, by proper design of the channels inside the head and the adjustment of the deposition parameters, a homogeneous film can be obtained. The smaller film shown here represents the first example presented in this work of simple direct ASD without the need of pre-patterning or etching steps.

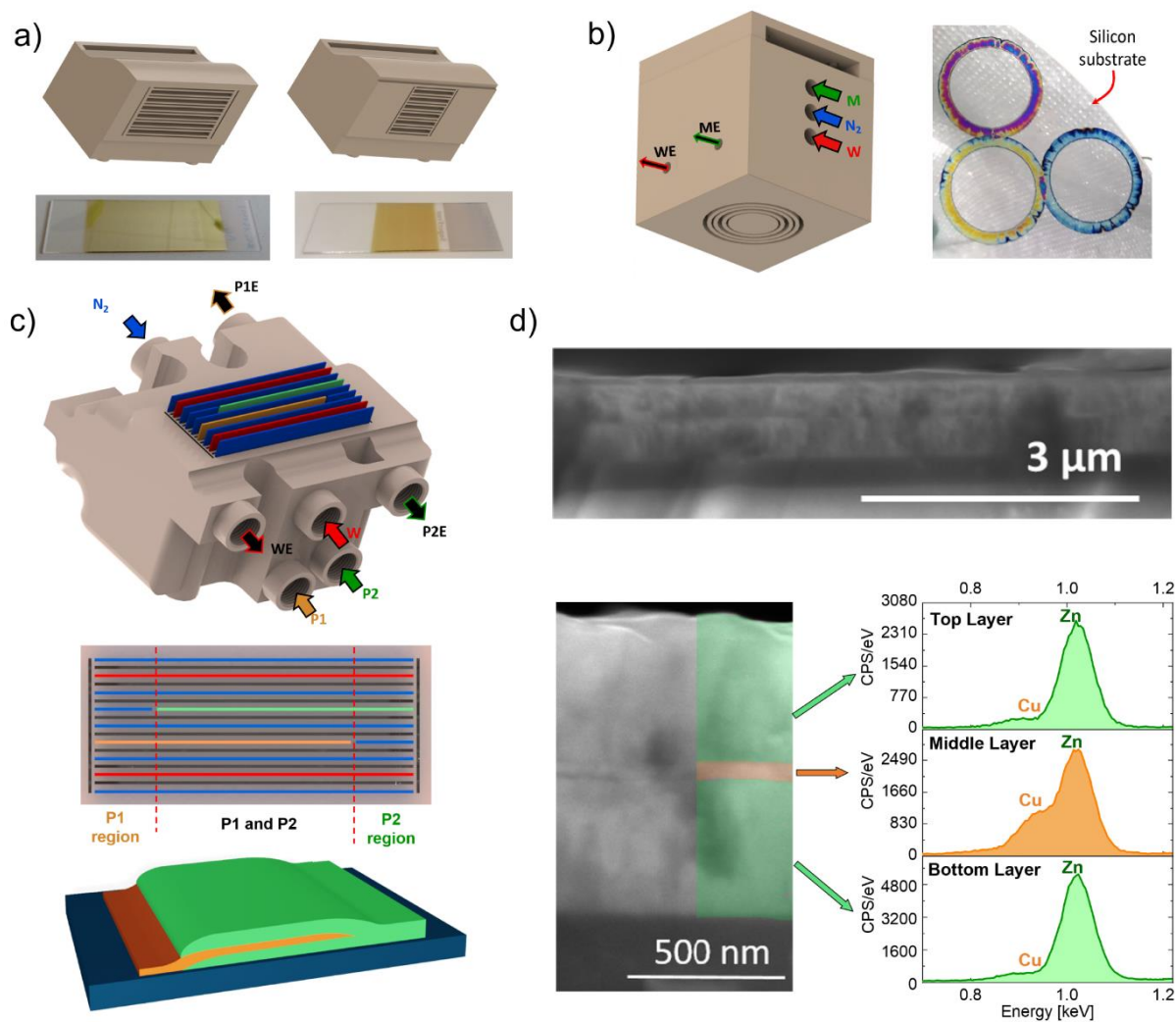

**Figure 3.** a) 3D scheme of two printed heads with different outlet lengths (5 cm vs. 2,5 cm). b) 3D scheme of a head designed for circular shape deposition in static SCVD mode. ZnO circles with different thicknesses are shown. c) 3D scheme of a customized head with two different metallic precursor outlets overlapping on the central part of the deposition area, and 3D view of the resulting multilayer pattern that can be obtained with such a head d) SEM pictures and corresponding EDS analyses of a ZnO/Cu$_2$O/ZnO stack deposited with the two-metal precursor head shown in c)

As stated above, the control of the gap between the head and the substrate is essential in order to ensure effective precursor separation and thus ALD reaction mode. But the possibility to mechanically adjust the gap is also interesting since it allows to tune the deposition regime between ALD and CVD (where precursors react in the gas phase above the substrate).[50] Deposition in spatial CVD (SCVD) mode may result convenient when dealing with flat, featureless substrates since the deposition rate is higher.[31,51] Also, SCVD has shown to be conformal when coating nanostructures.[25] Finally, the SCVD mode allows the use of precursors that would not react in the ALD mode (i.e. showing selective reaction with the substrate)[52]. In addition, we have shown in the past that the SCVD mode can be used to do selective deposition of lines when doing a static deposition (i.e. with an immobile substrate, since precursors can only meet in certain regions below the head if the deposition is static).[50] By combining this static deposition mode with the high flexibility in designing the injection head that 3D printing allows, SCVD heads can easily fabricated to deposit freeform patterns. Figure 3b shows the 3D scheme of a head in which the outlets have a circular shape instead of the straight outlets normally used in SALD. By doing static deposition in CVD mode with such a head, circular patterns can be directly printed, as shown in the figure. The thickness of the films can be controlled by the deposition time and flows used, yielding different colors.

Another interesting feature offered by 3D printing SALD deposition heads is the possibility of deposing different materials simultaneously in different locations, i.e. simultaneous patterning of different materials. This indeed opens the door to the direct deposition of devices containing different layers of materials in one run without any patterning steps once the head has been designed and printed and the process optimized. Figure 3c shows a 3D scheme of a head that has been designed to deposit two different materials, which overlap only in the central part of the deposition area. To do that, the head incorporates two independent metallic precursor outlets that serve to supply the two different precursors. These metal precursor outlets are shorter that the rest of the outlets (inert gas, oxidizer and exhaust), and are placed at opposite extremes of the head along the length of the outlets. The length of the outlet is completed with a short inert gas outlet. It is clear that such a head would be much more difficult, if not impossible at all, to fabricate by conventional machining, and that it would imply many more connections. 3D printing allows to have such an elaborate head in one block. The head has been tested to deposit a ZnO/Cu$_2$O/ZnO multilayer as the one shown in the scheme in Figure 3c. To do so, while ZnO was initially deposited, only nitrogen was injected through the copper precursor outlet. After a certain number of cycles were performed, then the zinc precursor was switched off and only nitrogen was injected through the outlet. Conversely, the copper precursor was switched on by starting the nitrogen bubbling through the precursor and $Cu_2O$ started depositing. The different layers of materials in the central part of the deposition where there is overlap can be clearly seen in the SEM cross-section image shown in Figure 3d. Because the deposition rate of $Cu_2O$ is much lower than ZnO (0.02 nm/s vs. 1 nm/s, respectively),[20,21] a thinner $Cu_2O$ layer was deposited for this demonstration, as clearly observed in the SEM close-up image in Figure 3d (~75 nm vs. ~475 nm). EDS analyses were performed on the different layers and, although the SEM probe is too big to have enough resolution, the results clearly show the presence of the different layers.

The examples presented above show that by simple design of customized heads and 3D printing of the same, selective deposition of functional materials with custom patterns can be readily achieved. Although this is already a significant step forward, allowing the direct selective deposition of functional materials, it requires a design and fabrication step for each case. And in all cases, the shape of the final deposition is going to be conditioned by the characteristics of the head, both when using it in SALD mode or SCVD mode. For certain applications, it would be desirable to be able to draw freeform patterns and to have the possibility of topological fabrication in the 3 dimensions. This indeed can again be done by proper design of SALD heads. While previous research on SALD has put the focus on the convenience of the technique towards up-scaling thanks to the atmospheric processing and high deposition rate, SALD heads can also be miniaturized, and this can be used to obtain selective deposition. **Figure 4**a shows the 3D scheme of what could be called a SALD pen. Figure 4b shows a SALD pen 3D printed in clear resin. The head is cylindrical and presents circular gas outlets arranged in a concentric way (Figure 4c). With such a head, freeform deposition can be achieved since in any direction that the head is moved the substrate is going to be alternatively exposed to the different precursors. The versatility offered by 3D printing is again very convenient in this case for the design and fabrication of the SALD pen shown. Such a head can then be easily implemented in a 3D table to perform the direct printing of functional materials with freeform patterns (Figure 4d).

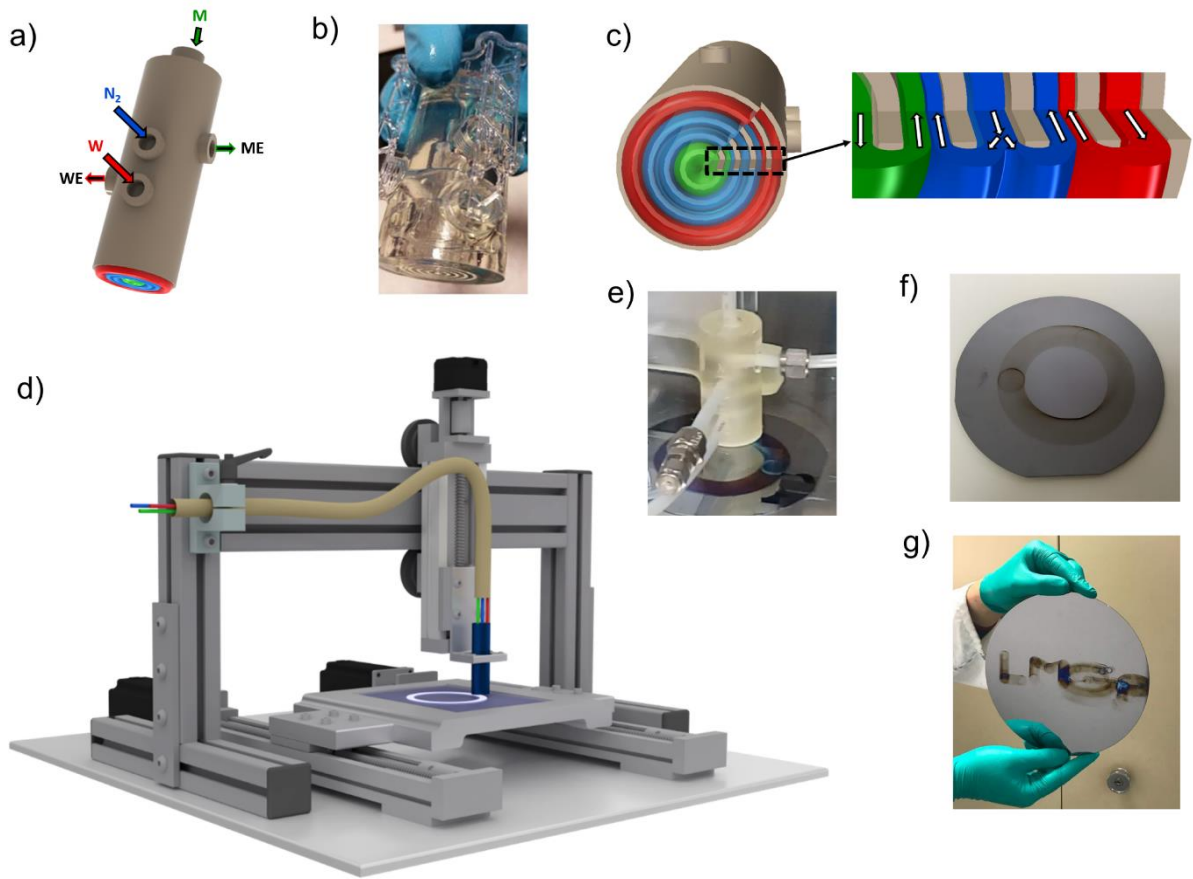

**Figure 4.** a) Scheme of a custom designed SALD pen. b) Picture of a SALD pen just after being 3D printed. c) View from the bottom of the SALD pen approach where the concentric circular gas outlets allowing deposition in any direction can be clearly seen. d) Scheme of the SALD pen installed in a 3D table. e) Image of the printed SALD pen installed in the 3D table and drawing a ZnO circle. f) ZnO circle and f) LMGP initials on a Si wafer drawn with the 3D printed SALD pen**.**

Although inkjet printing has similar characteristics that the printing approach proposed here, there are several assets that make our SALD based approach unique. In the first place, our approach is based on a gas phase deposition, thus being suitable for applications and processes where the use of solvents is not suitable or, simply, not possible. Secondly, the printing approach proposed is based on the ALD technique, and this implies that the resolution in Z can be controlled down to the sub-nanometer, something not possible with inkjet printing. Also, the approach benefits from the other unique assets of ALD, namely, conformality, density and compactness of the deposited films, and high material quality at low deposition temperature.

This is important when comparing with other deposition systems based on spray coating or plasma aerosol jets that are based on CVD reactions,[53] which yield less homogeneous thin films. Figure 4e shows an image of our 3D printed SALD pen installed in a 3D table, used in this case to directly print a circle of ZnO without any previous patterning step. Figure 4f shows a 50 nm thick ZnO circle on a silicon substrate printed with our SALD pen. Note that the lateral thickness of the circle is smaller than the diameter of the SALD pen. This is so since the reaction only takes place in the area below the pen where the metallic precursor is exposed (i.e. below the metal precursor outlet and adjacent exhaust outlet). The minimal lateral resolution of the system is thus defined by the design of the head, in particular the diameter of the outlet for the metallic precursor and the distance between the later and the adjacent exhaust outlet. Nevertheless, the lateral resolution can also be affected by the deposition parameters (i.e., whether the system is operated in SALD or SCVD mode). Another example of freeform direct printing with our SALD pen is shown in Figure 4g, were the initials of our lab (LMGP) have been written with a 120 nm thick ZnO film. While previous approaches use a miniaturized plasma gun to selectively modify the substrate prior to a conventional ALD deposition,[54] our approach allows the direct deposition of functional materials at low temperature with complex patterns in the open air. This is the first time that ALD, with all its assets, is implemented as an open-air 3D printing method, and the first time as well than a gas-based 3D printing approach is proposed.

## 3.- Conclusion

In summary, we present here a new approach for the 2D and 3D printing of functional materials, with nanometric resolution in Z, which is based on an open-air close-proximity SALD approach. By taking advantage of the design freedom that additive manufacturing offers, customized SALD heads can be easily and rapidly designed and printed, and at prices much lower than standard heads fabricated by traditional methods. With such customized heads, the area of the deposition can be easily adjusted and complex multimaterial devices can be printed without the need of patterning steps. By performing depositions in static SCVD mode, complex patterns can be easily made with custom printed heads. Finally, by reducing the SALD head in size and implementing concentric circular gas outlets, a SALD pen has been designed and printed that allows the open-air printing of functional materials. While the proof of concept presented here is limited to several mm in lateral resolution, a further miniaturization of the SALD pen through optimized engineering should allow a finer lateral resolution, down to the micron scale.[54,55] The same applies to the other ASD strategies proposed here. Our approach represents a breakthrough in the field of functional materials, opening the path to a simpler fabrication of complex and custom devices. For instance, the outlets can be designed to have a non-uniform shape so that non-homogenous films are grown on purpose so that combinatorial studies can be readily performed.[56]

**4.- Experimental Section**

*Design and printing of the SALD heads:* All the heads were designed with a commercial CAD software (SolidEdge). A stereolithography–based 3D printer (Form 2, from Formlabs) was employed. The printer uses the dimensional CAD-model to print the file in a layer per layer fashion, building up in the Z-axis direction. The printed heads were exposed 15-30 min to an UV light for post-treatment curing process as explained by the manufacturer. In this work only Clear resin from Formlabs is employed. However, it should be indicated that there are other alternative resins like High temperature or Tough that can be used depending on the operational conditions required during the SALD process. The metal heads were printed in aluminium by direct metal laser sintering (INITIAL, PRODWAYS GROUP).

*Deposition parameters:* ZnO films were deposited using Diethylzinc, (Aldrich, min. 95%) as metal precursor and deionized water as oxidizer, both used at room temperature. the substrate was heated up to 100 °C. $Cu_2O$ thin films were deposited using using hexafluoroacetylacetonate

CuI trimethylvinylsilane (hfac)CuI(tmvs), CupraSelect® as precursor, heated up to 65 °C and water as oxidized (at room temperature). $Cu_2O$ thin films in Figure 3a were deposited on glass slides (surface area of 2.6x7.6 cm²). The scanning speed was approximately 15 cm/s and the deposition's temperature was fixed at 240 °C. The films are about 80 nm thick. Regarding the static deposition of ZnO by SCVD with the head showed in Figure 3.b, the substrate was previously heated at 100 °C and the gap was set to ~150 µm. The deposition was done without moving the silicon substrate, previously cleaned using acetone, ethanol and isopropanol. The deposition times were 1'00", 1'30" and 2'00", for the blue, yellow and purple ring, respectively. The choice of the flow parameters and the gap was based on a previous study.[50] For the ZnO circle drawn with the SALD pen shown in figure 4e the number of cycles was fixed at 250 cycles – where one cycle corresponds to one complete circle deposition and the LMGP initials. For the LMGP initials, each letter was designed with a CAD software (SolidEdge) and then they were exported to an AutoCAD motion software in the form of the code language. Similarly, the temperature deposition was always fixed at 100 °C and the number of cycles was fixed at 300 cycles for each letter.

*Characterization:* The morphology of the thin films was characterized by scanning electron microscopy (SEM-FEG Environmental FEI QUANTA 250), EDS (SEM-FEG-ZEISS Ultra 55 equipped with an Edax CCD Hikari Pro Camera).

**Acknowledgements**

D.M.-R. acknowledges support from the European Union's Horizon 2020 FETOPEN-1-2016-2017 research and innovation program under Grant Agreement 801464, and through the Marie Curie Actions (FP7/2007-2013, Grant Agreement No. 63111). Funding from the University Grenoble Alpes is acknowledged through the AGIR POLE 2016 program under the grant SURPASS. The Agence Nationale de Recherche (ANR, France) is also acknowledged for funding via the programs ANR-16-CE05-0021 502 (DESPATCH). This project was also financially supported by ''Carnot Energies du Futur'' (ALDASH project) The Consejo Nacional de Ciencia y Tecnología (CONACYT, Mexico (No. 456558)), the ''ARC Energies Auvergne-Rhône Alpes'' and the French National Research Agency (in the framework of the ''Investissements d'avenir'' program (No. ANR-15-IDEX-02) through the project Eco-SESA) are acknowledged for PhD Grants. The authors acknowledge the CMTC characterization platform of Grenoble INP (Institute of Engineering) supported by the Centre of Excellence of Multifunctional Architectured Materials "CEMAM" n°ANR-10-LABX-44-01 funded by the "Investments for the Future" Program. Dominic de Barros is acknowledged for technical support.